\documentclass[12pt]{iopart}
\usepackage{graphicx}
\usepackage{amssymb}
\begin{document}
\title[The AWE Hypothesis: the Missing Link between DM and DE]{The Abnormally Weighting Energy Hypothesis:\\ the Missing Link between Dark Matter and Dark Energy}
\author{J.-M. Alimi$^{1,2}$, A. F\"uzfa$^{1,2}$}
\address{
$^1$ CNRS, Laboratoire Univers et Th\'eories (LUTh), UMR 8102 CNRS, Observatoire de Paris, \\
Universit\'e Paris Diderot ; 5 Place Jules Janssen, 92190 Meudon, France
\\
$^{2}$Groupe d'Application des MAth\'ematiques aux Sciences du COsmos (GAMASCO), 
University of Namur (FUNDP), Belgium}

\begin{abstract}
We generalize tensor-scalar theories of gravitation by the introduction of an abnormally weighting type of energy. This theory of tensor-scalar anomalous gravity is based on a relaxation of the weak equivalence principle
that is now restricted to ordinary visible matter only. As a consequence, the convergence mechanism toward general relativity is modified and produces naturally cosmic acceleration as an inescapable gravitational feedback induced by the mass-variation of some invisible sector. 
The cosmological implications of this new theoretical framework are studied. From the Hubble diagram cosmological test \textit{alone}, 
this theory provides an estimation of the amount of baryons and dark matter in the Universe that is consistent with the independent cosmological tests of Cosmic Microwave Background (CMB) and Big Bang Nucleosynthesis (BBN). Cosmic coincidence is naturally achieved from a equally natural assumption on the amplitude of the scalar coupling strength. Finally, from the adequacy to supernovae data, we derive a new intriguing relation between the space-time dependences of the gravitational coupling and the dark matter mass, providing an example of crucial constraint on microphysics from cosmology. This glimpses at an enticing new symmetry between the visible and invisible sectors, namely that the scalar charges of visible and invisible matter are exactly opposite.
\end{abstract}

\maketitle

\section{Introduction}
Recent observational evidence \cite{wmap1,wmap3,wmap5,perlmutter,riess,snls} indicates that the Universe is presently dominated by an intriguing component dubbed
Dark Energy (DE). Its gravitational action is to drive the current cosmic acceleration by mimicking a fluid of puzzling negative pressure acting on cosmological scales. The ultimate explanation of the physical origin of DE is often thought of
as the bridge between microphysics and gravitation. 
The widespread interpretation of DE based on the cosmological constant $\Lambda$ constitutes an acknowledged example of an intimate link between cosmology and particle physics.
In fact, $\Lambda$, introduced by Einstein himself as a Mach principle-inspired term \cite{einstein}, is currently interpreted as a non-vanishing vacuum energy. Although a huge and still unexplained fine-tuning is still required to reduce drastically the theoretical expectation of the cosmological constant value \cite{weinberg}, nevertheless, it enters the description of the dark sector within the so-called concordance model $\Lambda\textrm{CDM}$ together with the cold dark matter CDM. 
Furthermore, this model faces a triple coincidence problem: why do we live in an almost flat Universe ($\Omega_T =1$) with roughly the same amount of baryons, DM and DE today 
($\Omega_b=0.04\approx\Omega_{DM}=0.2\approx\Omega_{DE}=0.76$)? More specifically, how could the vacuum energy be precisely of the same order of magnitude of other present cosmological components? Instead, the measured amount of DE suggests that it is ruled by some cosmological mechanism such as quintessence or generalised additional fluid components \cite{copeland} whose origin has to be found in high-energy physics. However, one can expect \cite{brax} that the difficulties encountered in trying to overcome the coincidence issues and the related problems in high-energy physics and gravitation will remain as long as DE will be regarded as an additional component \textit{independent} of baryons and DM.
\\
\\
A possible way out of the coincidence problem could be in considering a more sophisticated physics for the whole dark sector \cite{amendola,farrar,corasaniti},
for instance by introducing in this sector new long-ranged interactions. But so far these works
have once again relied on negative pressures to explain cosmic acceleration. The interesting point however is that these novel interactions in the dark sector makes \textit{only} the mass of the invisible particles varying which constitutes a violation of \textit{weak equivalence principle} (WEP). This principle establishes the universality of free-fall for non-gravitationally bound objects, namely that gravitation stays unsensitive to their various nature and composition. In other words,
gravitation couples universally to non-gravitational energies (see also \cite{will} and references therein for an insightful presentation). This assertion has been extremely well-verified, notably at the $10^{-12}$ level with laboratory masses \cite{su}. However, these tests hold for ordinary matter \textit{only}, while the question of its validity to an invisible sector still remains an open debate, on both observational \cite{kamionkowski} and theoretical \cite{farrar,massd} points of view.
But if the WEP does not apply to an invisible sector, then
the \textit{strong equivalence principle} (SEP), that includes the WEP and extends it to gravitational binding energies, also does not hold anymore. It is indeed clear that the gravitational energy of mass-varying invisible matter particles \textit{do not weigh} in the same way than the gravitational energy of ordinary matter particles
with constant masses. Therefore, in a mixture of ordinary matter and mass-varying invisible matter, like the large-scale Universe, one should expect to observe an inescapable violation of the SEP. This crucial point has not been investigated in previous works on coupled DM-DE. In this paper, we will show that cosmic acceleration is precisely the observable trace of this SEP violation coming from the fact that the WEP does not apply to the invisible sector of cosmology.
\\
\\
Of course, the idea of a violation of the equivalence principle for the particular case of DM is not new and even appeared prior to the numerous evidence for cosmic acceleration and the advent of DE. Several models based on microphysics have been considered to achieve such a mass-variation for DM in particular (cf. \cite
{farrar,massd} and references therein) but the inescapable consequences of this WEP violation in the invisible sector on the SEP have 
not yet been explored. Therefore, the key point to be used to unveil the microscopic nature of gravitation indisputably rests on a critical discussion of \textit{both equivalence principles}, that would in the same time address the question of the physical nature of the dark sector. The results presented in this paper are precisely a way of achieving this double goal. \\
\\
In this paper, our basic assumption is that the invisible sector is constituted by an Abnormally Weighting type of Energy (AWE Hypothesis) \cite{awebi,awe,awedm,awedm2}, 
i.e. the WEP does not apply to the dark sector and has therefore to be relaxed.
Doing so, one must also consider the consequent violation of the SEP as we mentionned above, meaning that the usual laws of gravitation are modified and ordinary visible matter experiences a varying gravitational strength due to the existence of the invisible gravitational outlaw. 
In such a framework, it is clear that cosmic expansion is affected and we will show that the observed cosmic acceleration can be accounted for exclusively with this mechanism,
without requiring to any explicit negative pressures. 
This is why we will discard the cosmological constant all along this paper, leaving the question of its fine-tuning unexplained, in order
to show without confusion that the cosmological constant is not needed in the AWE framework to provide cosmic acceleration.
Moreover, this new explanation of DE in terms of the gravitational feedback of the invisible AWE will put a new light on four intricate \textit{theoretical}
problems of the cosmic concordance,
that we could dub the "AWE\textit{-full}" problems. 
\begin{itemize}
\item[(i)] The first of these problems is of course that of the nature of DM and the origin of DE, usually they are both assumed physically unrelated.\\
\item[(ii)] Strong negative pressures are unavoidable to provide acceleration in standard cosmology based on general relativity (GR): it is necessary to violate the strong energy condition \cite{caroll}: $p<-\rho/3$. However, these pressures could instead arise as an effective effect in matching a standard interpretation of cosmic expansion out of a modified gravity framework. In this latter case only, there will be other independent tests of the nature of DE than the difficult measurement of the equation of state $\omega=p/\rho$ and its cosmological evolution.
\\
\item[(iii)] The triple coincidence we face relies on several different physical processes: inflation (that
fixes $\Omega_T$ close to unity), baryogenesis and the production of DM (that respectively fixes $\Omega_b$ and $\Omega_{DM}$ to such a serendipitous and time-independent
ratio of order $5$) and the low-energy mechanism that makes DE the dominant component in the Universe today. The cosmological constant 
cannot explain this latter coincidence as this approach just introduces an arbitrary and absolute fundamental scale in cosmology. However, by relating cosmic acceleration
to matter, one can hope to get rid of the last of those coincidences on DE by reformulating it in terms of the second on the matter sector.
This is precisely what we will achieve here.\\
\item[(iv)] The fate of the Universe in the $\Lambda\textrm{CDM}$ scenario is the endless cosmic acceleration of a de Sitter Universe. Even worst, numerous cosmological data \cite{wmap5,snls} seem to favour a phantom DE, i.e. $\omega<-1$, for which the cosmic history achieves in the future into a Big Rip \cite{bigrip} singularity ($a\rightarrow\infty$ for some finite time). In the explanation presented here, the cosmic acceleration is transient albeit mimicking a phantom DE for some time.
\end{itemize}
We will come back to these AWE-full problems of cosmic concordance in the conclusion to show what are the possible solutions proposed by the AWE hypothesis.
\\
\\
The structure of this paper is as follows.
In section 2, we build from the AWE hypothesis a tensor-scalar theory of anomalous gravity
that naturally generalises the usual tensor-scalar theories (TST) of gravitation \cite{ts,bd,convts,barrow,damour2}. Indeed, while these TST \textit{only} considered a violation of the SEP, this new framework allows to adequately describe
the modifications of gravitation that are introduced by a relaxation of the WEP, and the consequent violation of SEP. The modified convergence mechanism toward general relativity with standard ordinary matter and an AWE component are studied in the case of general equations of state.
This extends our preceeding works with the particular cases of a Born-Infeld gauge interaction \cite{awebi,awe} and pressureless fluids in \cite{awedm}.
In section 3, we relate this modified convergence mechanism to cosmic acceleration. 
We also establish there the very general conditions upon which cosmic acceleration can be achieved.
We derive in section 4 remarkable cosmological predictions from the analysis of Hubble diagrams of far-away supernovae that unveils the nature of the abnormally weighting component and the origin of cosmic acceleration. Section 5 will be dedicated to a detailed discussion of the solution to the cosmic coincidence problem and the influence
of the parameters, deepening our previous work on this issue \cite{awedm}. We then use SNe Ia data in section 6 to constraint the scalar charges of this model
and show an interesting evidence for a new symmetry between the visible and hidden sectors of cosmology. 
In the conclusion, we review all the assumptions and main achievements of this paper, as well as the possible answers
brought to the AWE-full problems, before giving both theoretical and observational perspectives for further testing the AWE hypothesis.
\section{Cosmological dynamics in tensor-scalar anomalous gravity}
The \textit{Abnormally Weighting Energy} (AWE) hypothesis \cite{awebi,awe,awedm,awedm2} suggests an interpretation of DE in terms of a relaxation of the WEP, this last being now restricted to standard ordinary visible matter.
The physical framework contains three different sectors: gravitation, the common matter (baryons, photons, etc.) and the invisible sector, made of some \textit{abnormally weighting energy}.
Then, the laws of gravitation
have to rule both the expansion of space-time and the variation of the gravitational strengths that are experienced differently by the two matter sectors \cite{awedm, damour,catena}.
From the point of view of visible matter, these different gravitational strengths are not perceptible because the masses of ordinary matter particles are constants as a consequence of the restricted WEP.
However, the masses of invisible particles vary manifestly meanwhile the gravitational strength changes, yielding to cosmic acceleration as we shall see further. 
Therefore, the restriction of the WEP to ordinary matter can be observed through the violation of the SEP, i.e. the changing gravitational coupling that it yields. 
This can be expressed in the observable Dicke-Jordan frame by the following action ($c=1$): 
\begin{eqnarray}
\label{s_awe}
S&=&\frac{1}{16\pi G}\int\sqrt{-\tilde{g}}d^4\tilde{x}\left\{\Phi\tilde{R}-\frac{\omega_{BD}(M(\Phi))}{\Phi}\tilde{g}^{\mu\nu}\partial_\mu\Phi\partial_\nu \Phi\right\}+S_m\left[\psi_m, \tilde{g}_{\mu\nu}\right]\nonumber\\
&&+S_{awe}\left[\psi_{awe},M^2(\Phi)\tilde{g}_{\mu\nu}\right]
\end{eqnarray}
where $G$ is the "\textit{bare}" gravitational coupling constant (reducing to Newton's constant $G_N$ on Earth), $\tilde{g}_{\mu\nu}$ is the metric coupling universally to ordinary matter, $\Phi$ is a scalar degree-of-freedom scaling the observed gravitational strength $G_c=G/ \Phi$, $\tilde{R}$ is the scalar curvature build upon $\tilde{g}_{\mu\nu}$, $\omega_{BD}(\Phi)$ is the Brans-Dicke coupling function while $\psi_{m,awe}$
are the fundamental fields entering the physical description of the matter and abnormally weighting sectors, respectively. It is assumed here that the fields of the visible and invisible sectors $\psi_m$ and $\psi_{awe}$ do not couple directly (or extremely weakly) so that a direct observation of WEP violation on matter cannot be observed from this channel in any precision test of the WEP.
In this model, the local laws of physics for ordinary matter are those of special relativity (as the matter action $S_m$ does not explicitely depend on the 
scalar field $\Phi$) while the abnormally weighting dark sector $S_{awe}$ exhibits a mass-variation (represented by the non-minimal coupling 
$M(\Phi)$). The action (\ref{s_awe}) therefore generalizes usual TST \cite{ts,bd,convts,barrow} which consider a violation of the SEP only by encompassing the whole physics of the \textit{equivalence principles} (EPs) violation due to the anomalous gravity of the dark sector. In the following, a $\tilde{}$ will denote observable quantities, i.e. quantities expressed in the Dicke-Jordan frame.\\
\\
Although the action (\ref{s_awe}) describes correctly what can be observed in the presence of an abnormally weighting sector, this $Dicke-Jordan$ observable frame 
somehow hide the underlying theory of gravitation due to the admixture of scalar and tensor degrees of freedom and is consequently not convenient for studying
the dynamics of the gravitational coupling $\Phi$ \cite{convts}. It is therefore insightful 
to rewrite action (\ref{s_awe}) in the so-called
Einstein frame where the tensorial $\tilde{g}_{\mu\nu}$ and scalar $\Phi$ gravitational degrees of freedom separate into a bare metric $g_{\mu\nu}$ and a screening field $\varphi$. Moving between
the Dicke-Jordan observable frame and the Einstein frame is effected by performing the conformal transformation :
the Dicke-Jordan observable frame and the Einstein frame is effected by performing the conformal transformation :
$$
\tilde{g}_{\mu\nu}=A_m^2(\varphi)g_{\mu\nu}
$$
together with the identifications:
\begin{eqnarray}
G/G_c&=&\Phi=A_m^{-2}(\varphi),\label{gc}\\
3+2\omega_{BD}&=&\left(\frac{d\ln A_m(\varphi)}{d\varphi}\right)^{-2}\nonumber\\
M(\Phi)&=&\frac{A_{awe}(\varphi)}{A_m(\varphi)}\cdot\nonumber
\end{eqnarray}
Doing so, the action (\ref{s_awe}) can be rewritten
\begin{eqnarray}
\label{s_awe2}
S&=&\frac{1}{16\pi G}\int\sqrt{-g}d^4x\left\{R-2g^{\mu\nu}\partial_\mu\varphi\partial_\nu\varphi\right\}+S_m\left[\psi_m,A_m^2(\varphi)g_{\mu\nu}\right]\nonumber\\
&+&S_{awe}\left[\psi_{awe},A_{awe}^2(\varphi)g_{\mu\nu}\right]\cdot
\end{eqnarray}
It should be reminded that the observable quantities are not directly obtained in this frame as the physical units are universally scaled here with $A_m(\varphi)$ (the standards
of physical units like meter and second are defined with rods and clocks that are build upon the matter fields $\psi_m$). In particular, the inertial masses of the ordinary and abnormally weighting matter sectors are scaled respectively with $A_m(\varphi)$
and $A_{awe}(\varphi)$ in the Einstein frame, while the Planck mass defines the bare gravitational strength $\bar{m}_{Pl}=G^{-2}$. The AWE hypothesis assumes that the invisible sector experiences background space-time ($g_{\mu\nu}$) with a different gravitational strength than ordinary visible matter, which is formulated
in terms of the nonuniversality of the nonminimal couplings $A_m(\varphi)\ne A_{awe}(\varphi)$, an idea echoing the effective theories of gravitation derived from string theory \cite{damour2,gasperini1,gasperini2}. This nonuniversality of gravitation is therefore a \textit{minimal} violation of the WEP taken as a whole: there still exists one universal background space-time
and its intrinsic geometric properties (the light cone for instance) but it is experienced through two different varying gravitational couplings.
\\
\\
The Einstein field equations that derive from the action (\ref{s_awe2}) can be written down
\begin{eqnarray}
\label{einstein}
R_{\mu\nu}-\frac{1}{2}R g_{\mu\nu}&=&8\pi G\left(T_{\mu\nu}^{m}+T_{\mu\nu}^{\textrm{awe}}\right)+2\partial_\mu\varphi\partial_\nu\varphi-g_{\mu\nu}g^{\alpha\beta}\partial_\alpha\varphi\partial_\beta\varphi
\end{eqnarray}
where $T_{\mu\nu}^{i}=\frac{2}{\sqrt{-g}}\frac{\partial \mathcal{L}_i \sqrt{-g}}{\partial g^{\mu\nu}}$ are the stress-energy tensors for the sector $i$. 
The Klein-Gordon equation for the scalar field is
\begin{eqnarray}
\label{kg}
\Box\varphi=-4\pi G \left(\alpha_m(\varphi)T^(m)+\alpha_{awe}(\varphi)T^{(awe)}\right)
\end{eqnarray}
where $T^(i)$ is the trace of stress-energy tensor of sector $i$, $\Box\varphi=g^{\mu\nu}\nabla_\mu\nabla_\nu\varphi$ and $\alpha_i(\varphi)=\frac{d\ln A_i(\varphi)}{d\varphi}$ the scalar coupling strengths
to the ordinary and abnormally weighting sectors, respectively. In the absence of direct couplings between $\psi_m$ and $\psi_{awe}$,
both sectors are conserved separately, and their conservation equations in Einstein frame are given by
\begin{equation}
\label{cons}
\nabla_\mu T_\nu^{\mu}(i)=\alpha_i(\varphi)T^{(i)}\partial_\nu\varphi
\end{equation}
for the $i$th sector.
\\
\\
The AWE hypothesis has to be consistent with the stringent constraints on the equivalence principles \cite{will}.
This is indeed the case. Firstly, the masses of the ordinary matter particles are ruled by a restricted WEP and therefore do not vary (in the observable frame of Eq.(\ref{s_awe})). Consequently, the non-gravitational laws of physics of ordinary matter are not modified.
However, as we shall see in the next section, their gravitational interactions are modified, because the gravitational binding energies vary according to the amount of AWE relative to that of ordinary matter. One can therefore expect these modifications of gravity to be extremely weak on small scales as they are limited by the locally very faint abundance of invisible matter. However, on cosmological scales, invisible matter is profuse and substantial departures from GR are expected. 
\\
\\
Let us therefore study the cosmological evolution of the observed large-scale gravitational strength $G_c$ (\ref{gc}) (see also \cite{awedm}). Assuming 
a flat Friedmann-Lema\^itre-Robertson-Walker (FLRW) background space-time $g_{\mu\nu}$ and a fluid description of the matter and AWE sectors $S_m$ and $S_{awe}$, we obtain 
from Eqs. (\ref{einstein}-\ref{cons}) the Friedmann, acceleration and scalar field equation
\begin{eqnarray}
\label{friedmann}
\left(\frac{\dot{a}}{a}\right)^2=\frac{\dot{\varphi}^2}{3}+\frac{8\pi G}{3}\left(\rho_m+\rho_{awe}\right)\\
\frac{\ddot{a}}{a}=-\frac{2}{3}\dot{\varphi}^2-\frac{4\pi G}{3}\left[\left(\rho_m+3p_m\right)+\left(\rho_{awe}+3p_{awe}\right)\right],\label{acc}\\
\ddot{\varphi}+3\frac{\dot{a}}{a}\dot{\varphi}=-4\pi G\left\{\alpha_m(\varphi)\left(\rho_m-3p_m\right)+\alpha_{awe}(\varphi)\left(\rho_{awe}-3p_{awe}\right)\right\}\label{kg_awe}
\end{eqnarray}
where a dot denotes a derivation with respect to the time $t$ in the Einstein frame and where $\rho_X$ and $p_X$ stand for the energy density and 
pressure of the fluid X ($\equiv$ matter or AWE) in the Einstein frame. These \textit{bare} energy densities are related to the observable ones $\tilde{\rho}_X$ by $\rho_X=A_m^4(\varphi)\tilde{\rho}_X$. Similar relation holds for the pressures.
In the Einstein frame, it is important to notice that there cannot be any acceleration unless one of the fluid violates of the strong energy condition so that $p_i<-\rho_i/3$. 
The bare expansion $a$ is then always decelerated $\ddot{a}<0$ in this frame while cosmic acceleration might occur in the observable frame under appropriate conditions on 
the scalar field dynamics. The conservation equations (\ref{cons}) now become
\begin{eqnarray}
\nonumber
\dot{\rho}_{i}+3\frac{\dot{a}}{a}\left(\rho_{i}+p_{i}\right)=\alpha_{i}(\varphi)\; \dot{\varphi}\left(\rho_{i}-3p_{i}\right)
\end{eqnarray}
and can be directly integrated to give
\begin{eqnarray}
\label{rhos}
\rho_{X} &=& A_{X}^{1-3\omega_{X}}(\varphi) \rho_{X,i}a^{-3(1+\omega_{X})}
\end{eqnarray}
where $\omega_{X}=p_{X}/\rho_{X}$ is the equation of state for the fluid $X$.
In the above equation, $\rho_{X,i}$ are arbitrary constants and
we will denote in the following by $R_i$ the ratio of $\rho_{m,i}$ over $\rho_{awe,i}$ at a given epoch $i$.\\
\\
Let us now reduce (\ref{kg_awe}) to a decoupled equation by using the number of e-foldings : $\lambda=\ln (a/a^i)$ as a time variable. Using (\ref{friedmann}) and (\ref{acc}), the Klein-Gordon equation (\ref{kg_awe}) now reduces to
\begin{equation}\label{dyn_1}
\frac{2\varphi''}{3-\varphi^{'2}}+(1-\omega_T)\varphi'
+\alpha_m(\varphi)(1-3\omega_T)+\frac{(\alpha_{awe}(\varphi)-\alpha_m(\varphi))}{1+\frac{\rho_m}{\rho_{awe}}}(1-3\omega_{awe})=0
\end{equation}
where a prime denotes a derivative with respect to $\lambda=\ln (a/a^i)$ and where the total equation of state $\omega_T$ for the admixture of matter and AWE fluids is $\omega_T=(p_m+p_{awe})/(\rho_m+\rho_{awe})$. The first three terms of Eq. (\ref{dyn_1}) represents the usual cosmological dynamics of TST with WEP \cite{convts,barrow} while the last brings all the information on the WEP relaxation. Obviously, the WEP can be retrieved and the well-known convergence mechanism of usual TST holds in three different cases:
(i) the nonminimal coupling is universal $\alpha_m=\alpha_{awe}$ ($M(\Phi)=1$), (ii) the AWE fluid is relativistic $\omega_{awe}=1/3$ and (iii) the AWE sector is sub-dominant $\rho_m\gg\rho_{awe}$ ($M(\Phi)\rightarrow 1$). This last case occurs for instance during the radiative era when $\rho_m\sim a^{-4}$ while $\rho_{awe}\sim a^{-3}$
for non-relativistic AWE fluid.
In conclusion, the WEP can then be well approximated anywhere the amount of invisible AWE matter is negligible.
\\
\\
When the equations of state of matter and AWE are identical $\omega_m=\omega_{awe}=\omega_T=\omega$, Eq. (\ref{dyn_1}) can easily be reduced to an autonomous equation
\begin{equation}\label{dyn_2}
\frac{2\varphi''}{3-\varphi^{'2}}+(1-\omega)\varphi'+(1-3\omega)\aleph(\varphi)=0,
\end{equation}
with $\aleph(\varphi)=d\ln\mathcal{A}(\varphi)/d\varphi$ with $\mathcal{A}(\varphi)$ the coupling function resulting from the mixing of matter and AWE fluids:
\begin{equation}
\label{biga_w}
\mathcal{A}(\varphi)=A_{awe} (\varphi)\left[R_i^{-1}+\left(\frac{A_{m}(\varphi)}{A_{awe}(\varphi)}\right)^{1-3\omega}\right]^{1/(1-3\omega)}\cdot
\end{equation}
The above equations (\ref{dyn_2})-(\ref{biga_w}) generalize
the results presented in \cite{awedm} for matter-dominated era \textit{only}.
Indeed, during the matter-dominated era ($\omega=0$), we found in \cite{awedm}
\begin{equation}
\label{biga}
\mathcal{A}(\varphi)=A_m(\varphi)+R_i^{-1}A_{awe}(\varphi),
\end{equation}
and 
\begin{equation}
\label{aleph}
\aleph(\varphi)=\alpha_m(\varphi)+\frac{(\alpha_{awe}(\varphi)-\alpha_m(\varphi))}{1+R_i\frac{A_m(\varphi)}{A_{awe}(\varphi)}}\cdot
\end{equation}
Eq. (\ref{dyn_2}) is completely
analogous to that of a damped oscillator with a variable mass rolling down some potential given by the logarithmic derivative of the resulting coupling function $\mathcal{A}(\varphi)$. Therefore, the convergence mechanism of TST with WEP is preserved despite the violation of WEP. However, this mechanism depends on the relative
concentrations of ordinary matter and AWE. Indeed, the attracting value to which the scalar field is driven by the contest between matter and AWE is given by 
\begin{equation}
\label{phi_inf}
\alpha_m(\varphi_\infty)\frac{\rho_m(\varphi_\infty)}{\rho_{awe}(\varphi_\infty)}+\alpha_{awe}(\varphi_\infty)=0,
\end{equation}
with $\frac{\rho_m(\varphi)}{\rho_{awe}(\varphi)}=R_i\frac{A_m(\varphi)}{A_{awe}(\varphi)}$. The attracting value $\varphi_\infty$ depends
on the ratio of ordinary matter over AWE component and is intermediate between the 
value of $\varphi$ for which $A_m(\varphi)$ is extremum (when $\rho_m\gg\rho_{awe}$) and the value of
$\varphi$ for which $A_{awe}(\varphi)$ is extremum (when $\rho_m\ll\rho_{awe}$).
This density dependance yields a chameleon effect \cite{awedm, chameleon,kaloper}, i.e. that the effective mass of the field varies with its neighborhood. Furthermore, if 
visible and invisible matter undergo different gravitational collapses due, for instance, to the dissipative processes that would affect exclusively the first, the whole mechanism becomes scale-dependent and allows to retrieve locally GR, in a neighborhood where visible matter is dominant. This occurs notably in the highly visible matter-dominated environment of the Solar System. 
In consequence, substantial variations of $G_c$ (\ref{gc}) are expected on cosmological scales where the invisible component is profuse while on very low scales they will be quite low due
to the very high density ratio of visible over invisible matter.
\section{Cosmic acceleration emerging naturally from tensor-scalar anomalous gravity}
We study here the modifications brought by the AWE component on the convergence mechanism toward general relativity (GR) and how this revised mechanism
constitutes a natural process of the observed cosmic acceleration. Indeed, this acceleration can only be described in the Dicke-Jordan frame of Eq.(\ref{s_awe}), where
cosmic expansion is described by the scale factor $\tilde{a}=A_m(\varphi)a=a/\Phi^{1/2}$ that is measured with visible matter. The Hubble diagram test allows to measure it through
the luminous distance $d_L(\tilde{z})=(1+\tilde{z})\tilde{H}_0\int_0^{\tilde{z}}
d\tilde{z}/\tilde{H}(\tilde{z})$ where the observable Hubble parameter $\tilde{H}(\tilde{z})$ is given by
\begin{equation}
\label{hubble1}
\tilde{H}^2=\frac{8\pi G A_m^2(\varphi)}{3}\left(\tilde{\rho}_m+\tilde{\rho}_{awe}\right)\frac{3\left(1+\alpha_m\varphi'\right)^2}{3-\varphi^{'2}}\cdot
\end{equation}
The Friedmann equation (\ref{hubble1}) in the observable frame allows us to define the density parameters
of ordinary and AWE matter by 
\begin{equation}
\Omega_{m,\textrm{awe}}=\frac{8\pi G A_m^2(\varphi) \tilde{\rho}_{m,awe}}{(3\tilde{H^2})}
\end{equation} 
and $\Omega_\varphi=1-\Omega_m-\Omega_{\textrm{awe}}$.
The acceleration equation in the Dicke-Jordan frame is given by
\begin{eqnarray}
\label{acc_obs}
\frac{1}{\tilde{a}}\frac{d^2\tilde{a}}{d\tilde{t}^2}&=&\overbrace{\tilde{H}^2\frac{\varphi^{'2}\left(3\frac{d\alpha_m}{d\varphi}-2\right)-6\alpha_m\varphi'}{3\left(1+\alpha_m\varphi'\right)^2}}^{I}\overbrace{-\frac{4\pi G A_m^2(\varphi)}{3}\left(1+3\alpha_m^2\right)\tilde{\rho}_m}^{II}\nonumber\\
&& \underbrace{-\frac{4\pi G A_m^2(\varphi)}{3}\left(1+3\alpha_m\alpha_{awe}\right)\tilde{\rho}_{awe}\cdot}_{III}
\end{eqnarray}
In the above equation, there are only two terms that can possibly lead to cosmic acceleration: the term I, related to the dynamics of the scalar field, and the term III, in the case where the product $\alpha_{awe}\alpha_m$ would be negative. This term comprises the exchange of scalar particles between visible matter and AWE
and is actually the main contribution to cosmic acceleration (see Figure 1). 
In this figure, we have plotted separately the terms I, II and III (divided by $\tilde{H}^2$) together with their sum that
gives the observable acceleration factor $q=d^2\tilde{a}/d\tilde{t}^2/(\tilde{a}\tilde{H}^2)$ accounting for SNLS supernovae data \cite{snls}. From these data, it appears that only the term III is positive and is the main contribution to $q$ at late times. Cosmic acceleration requires $\alpha_{awe}\alpha_m<0$ or, equivalently, that
the scalar coupling strengths $\alpha_i$'s in Eq.(\ref{aleph}) have to be of opposite signs. In other words, the matter and AWE coupling functions should be inversely proportional:
\begin{equation}
\label{as}
A_{awe}(\varphi)=A_{m}^{-R_\infty}(\varphi)
\end{equation}
where $R_\infty=\rho_b/\rho_{awe}(t\rightarrow\infty)$ is the ratio at which baryons and AWE densities freeze once the scalar field reaches the attractor $\varphi_\infty$.
This parameter $R_\infty$ also measures the (absolute value of the) ratio between the opposite scalar charges of ordinary matter (baryons) and AWE related to the new interaction mediated by $\varphi$. 
\\
\\
Let us now examine what these general prescriptions yield on the resulting coupling function $\mathcal{A}(\varphi)$ (\ref{biga}).
By replacing (\ref{as}) into (\ref{aleph}), we find
$$
\aleph(\varphi)=\alpha_m(\varphi)\frac{R_i A_m^{1+R_\infty}-R_\infty}{R_i A_m^{1+R_\infty}+1}
$$
which cannot be singular as $A_m(\varphi)>0$ and admits two roots, either $\alpha_m(\varphi_0)=0$ or $A_m(\varphi_0)=\left(R_\infty/R_i\right)^{1/(1+R_\infty)}$, where the scalar field driving force $\aleph(\varphi)$ vanishes in (\ref{dyn_2}) (with $\omega=0$). To characterize these two fixed points, we have to estimate the sign of the quantity $d\aleph(\varphi)/d\varphi$.
For $\alpha_m(\varphi_0)=0$ ($A_m(\varphi_0)=1$ without loss of generality), we have 
\begin{equation}
\frac{d\aleph(\varphi)}{d\varphi}=\frac{d\alpha_m(\varphi)}{d\varphi}\frac{R_i-R_\infty}{1+R_i},
\label{dalephdphi}
\end{equation}
whose sign depends on various parameters. For the other fixed point, $A_m(\varphi_0)=\left(R_\infty/R_i\right)^{1/(1+R_\infty)}$, we find
\begin{equation}
\frac{d\aleph(\varphi)}{d\varphi}=\alpha^2_m(\varphi)R_\infty>0,
\end{equation}
which implies that this fixed point is an attractor\footnote{Provided there is a solution to $A_m(\varphi_0)=\left(R_\infty/R_i\right)^{1/(1+R_\infty)}$.}.
Therefore, in order to ensure an evolution departing from GR that could provide cosmic acceleration, we are lead to require that (i) $A_m(\varphi)$ admits an extremum $\varphi_0$ (so that $\alpha_m(\varphi_0)=0$) \textit{and} (ii) that this extremum is turned into a maximum of $\mathcal{A}(\varphi)$ (so that $d\aleph(\varphi)/d\varphi|_{\varphi_0}<0$ in (\ref{dalephdphi})).
Doing so, $\varphi_0$ will be an unstable fixed point.
This can be done by choosing $d\alpha_m(\varphi)/d\varphi|_{\varphi_0}>0$ (resp. $<0$) when $R_i<R_\infty$ (resp. $R_i>R_\infty$).
However, it is compulsory to assume a matter coupling function $A_m(\varphi)$ with a global minimum (i.e., $d\alpha_m(\varphi)/d\varphi|_{\varphi_0}>0$) if we want
to retrieve the usual convergence mechanism of TST when AWE is sub-dominant
and avoid the non-perturbative effects that leads to spontaneous scalarisation of compact objects \cite{gef2}. In conclusion, this predicts that we will find $R_i<R_\infty$ in reproducing cosmic acceleration with this mechanism.\\
\\
According to this general discussion, the shape of the effective potential Eq.(\ref{biga}) during the matter-dominated era on which the scalar field rolls in Eq. (\ref{dyn_2}) looks like the shape of mexican hat illustrated in Figure 2a. It has one unstable equilibrium at $\varphi=0$ corresponding to GR with bare gravitational strength $G$ and another stable equilibrium $\varphi_\infty$ where the theory is similar to GR but with a greater value of the observed gravitational strength $G_c=G A^2_m(\varphi_\infty)$ (\ref{gc}). The gravitational coupling starts from rest at the CMB epoch near the bare value corresponding to GR, unstable in the matter-dominated era, and accelerates downward the minimum on the right.
This accelerated growth makes background space-time expansion appearing stronger and stronger, therefore reproducing DE effect. When the coupling settles into the miminum, gravitation on large-scales resembles to GR but with a stronger coupling. This mechanism depends on the relative concentration of baryons and AWE, as shown here by the values of the parameter $R_i=\rho_b/\rho_{awe}$ at CMB, and fails to depart from GR when the amount of AWE is negligible ($R_i>1$). On Figure 2b, the reader will find the modification
of the shape of the effective coupling function $\mathcal{A}(\varphi)$ Eq. (\ref{biga_w}) that follows from a variation of the equation of state parameter $\omega$ shared
by the ordinary matter and AWE fluids. Starting from $\omega=-1$ where the shape of the effective coupling function $\mathcal{A}(\varphi)$ for the admixture is close to that of the matter one $A_m(\varphi)$, the mexican hat shape progressively settles when $\omega$ increases. When the fluids are relativistic $\omega=1/3$, the effective coupling function becomes singular, as these fluids decouple from the scalar field in Eq.(\ref{dyn_2}). Finally, for ultra-relativistic fluid $\omega>1/3$, the shape of the effective coupling function $\mathcal{A}(\varphi)$ approaches the one of $A_{awe}(\varphi)$ that has no global minimum and forces a departure from GR that can only be stopped by a decrease of the equation of state. This changes in the effective coupling function are particularly promising for building an inflation mechanism
where ordinary matter and AWE are constituted, for instance, by massive scalar fields.
However, this topic goes well beyond the scope of the present work and is left for future studies.
\section{Observational predictions of tensor-scalar anomalous gravity}
Challenging the cosmological observations with our new theoretical framework will allow us to unveil the nature of the AWE component.
To this end, we need to specify the coupling functions to be used in computing the scalar field dynamics and the cosmological observables.
From the above discussion in section 3, we need the coupling functions to AWE and matter to be reciprocal and a coupling function to matter with a global minimum.
This ensures both the existence of GR-like attractors in the modified cosmological convergence mechanism while staying compatible with tests of the SEP \cite{convts,gef}.
Without loss of generality, we can consider the following well-known parameterization for the couplings $\alpha_{m,awe} (\varphi)$ \cite{convts,awedm,gef2,gef}:
\begin{eqnarray}
\left\{
\begin{tabular}{l}
$\alpha_m(\varphi)=k\varphi$\\
\\
$\alpha_{awe}(\varphi)=-R_\infty k\varphi$\\
\end{tabular}
\right. 
\label{couplings}
\end{eqnarray}
where $k$ is the coupling strength to the gravitational scalar. A natural assumption of nonminimally coupled theories of gravitation is to consider this strength $k$ of order unity, namely that the coupling to scalar gravitational mode $\varphi$ has the same amplitude than those of the tensorial ones $g_{\mu\nu}$.
Doing so, there are three free parameters
in this model: (i) the relative amount of matter and AWE at start $R_i=\rho_m/\rho_{awe}(a_i)$), (ii) the parameter $R_\infty$ in Eq.(\ref{as}) and (iii) the value of the scalar field at start, $\varphi_i$, which illustrates the departure from GR at the end of the radiative era.
From there, the cosmic contest between AWE and ordinary matter for ruling the EP begins and the cosmological value of $G_{c}$ is slowly pushed back from GR, provided it did not start rigorously from GR value, before falling toward the attractor whose position depends on the relative concentrations of baryons and AWE (see also Figure 2). As long as $G_{c}$ (\ref{gc}) increases, ordinary matter
couples more and more strongly to the background expansion which therefore appears accelerating to visible matter observers. \\
\\
In order to proof the adequacy of the AWE hypothesis, we have performed statistical analysis of Hubble diagram data of far-away supernovae \cite{riess,snls}.
This model predicts the following density parameters for ordinary matter and AWE as well as for the age of the Universe : $\Omega_m=0.05^{+0.13}_{-0.04}$ 
($\Omega_{m}=0.06^{+0.08}_{-0.05}$), $\Omega_{\textrm{awe}}=0.22^{+0.11}_{-0.05}$ ($\Omega_{\textrm{awe}}=0.20^{+0.08}_{-0.06}$), 
($\Omega_{\varphi}=1-\Omega_{m}-\Omega_{\textrm{awe}}$)
and $t_0=13.1^{+0.6}_{-0.4}Gyr$ ($t_0=14.1^{+1.02}_{-0.55}Gyr$) at the 95\% confidence level for HST (SNLS) data sample ($H_0=70km/s/Mpc$ for the estimation of $t_0$). 
These results strongly suggest to identify ordinary matter to baryons, AWE to DM and the dynamics of the scalar field $\varphi$ to a DE component.
Doing so, the predicted amount of baryons is consistent with Big Bang Nucleosynthesis (BBN) and CMB constraints \cite{wmap1,BBN}.
These cosmological predictions look remarkably similar to the predictions of the concordance model $\Lambda$\textrm{CDM}
although they are obtained from the Hubble diagram \textit{alone}, i.e. from the homogeneous dynamics of the Universe, which 
is an outstanding achievement of the AWE hypothesis. 
Indeed, the $\Lambda\textrm{CDM}$ is only able to predict the total amount of pressureless matter 
($\Omega_{b}+\Omega_{DM}$) from the same cosmological test and requires a cross-analysis, for instance of the CMB anisotropies, to break this degeneracy. 
In the present case, the Hubble diagram becomes a fully independent test whose results are, in addition, consistent with WMAP constraints \cite{wmap1,wmap3}. Figure 3 illustrates the confidence regions in the plane ($\Omega_{b}$, $\Omega_{awe}\equiv\Omega_{DM}$) for the different parameterizations
of the coupling function ((\ref{couplings}) and (\ref{couplings2}) below). As well, we can measure, with supernovae data alone, the relative amount of baryons and AWE (i.e., DM according to the above identifications) at the beginning of the matter-dominated era (represented by the ratio of their densities at that time $R_i=\rho_b/\rho_{DM}$(CMB)).
We find $R_i=0.21^{+0.09}_{-0.07}$ ($R_i=0.22^{+0.13}_{-0.07}$) at the 68\% confidence level for HST (SNLS) data sample. 
By comparing the initial value of this ratio to the present one, we can deduce that the DM mass has decreased of approximately 30\% (cf. \cite{awedm}, Figure 9). 
\\
\\
In Figure 4,
we give a comparison between the observable scale factors accounting for supernovae data predicted by the present model specified by (\ref{hubble1}), (\ref{acc_obs}) and (\ref{couplings}) and the cosmological constant. Both cosmic expansions remains almost undistinguishable until the Universe reaches about twice its present size. Then, while the cosmological constant endlessly dominates
the Universe and drives an eternal exponential expansion, the AWE hypothesis suggests that baryons and AWE enter the final stage of their contest to rule the EP. This ultimate episode begins with the fall of the gravitational coupling into the attractor of Figure 2 which makes the scale factor oscillating. An equilibrium asymptotically establishes where baryons and DM do equally weigh (for $R_\infty =1$) and gravitation on large-scales is correctly described by GR with a different value of the coupling.
\\
\\
It is worth emphasizing that these results do not depend upon the particular parameterization Eq.(\ref{couplings}). Indeed, if we use another set of coupling functions, for instance,
\begin{eqnarray}
\left\{
\begin{tabular}{l}
$A_m(\varphi)=\cosh (k\varphi)$\\
\\
$A_{awe}(\varphi)=\cosh^{-R_\infty}(k\varphi)$\\
\end{tabular}
\right. \cdot
\label{couplings2}
\end{eqnarray}
which leads to a shape of effective coupling function $\mathcal{A}(\varphi)$ similar to the one of Figure 2a, then we find similar confidence regions (see figure 3b). The confidence regions have been computed in scanning the 3-D phase space of free parameters $(R_i, \; R_\infty, \; \varphi_i)$ with more than $2\times 10^7$ values, each representing a complete cosmological evolution. 
\section{An escape to the cosmic coincidence problem}
Let us now focus on how cosmic coincidence can be justified in this mechanism.
We can first interprete the observed cosmic acceleration described by (\ref{hubble1}) and (\ref{acc_obs}) as caused by a fictive exotic fluid with negative pressures
in Dicke-Jordan frame.
Following \cite{sahni}, let us match the Friedmann and acceleration equations (\ref{hubble1}) and (\ref{acc_obs}) in the Dicke-Jordan frame to
\begin{eqnarray}
\tilde{H}^2&=&\frac{8\pi G A_m^2}{3}\left(\tilde{\rho}_m\left(1+R_i^{-1}\right)+\tilde{\rho}_{DE}\right)\\
\frac{1}{\tilde{a}}\frac{d^2\tilde{a}}{d\tilde{t}^2}&=&-\frac{4\pi G A_m^2}{3}\left\{\tilde{\rho}_m\left(1+R_i^{-1}\right)+\tilde{\rho}_{DE}\left(1+3\omega_{e}\right)\right\}
\end{eqnarray}
where $\rho_{DE}$ and $\omega_{e}$ stand for the energy density and the equation of state of the fictive DE fluid. 
The fictive DE energy density can then be written down
\begin{equation}
\label{eff}
\tilde{\rho}_{DE}=\tilde{\rho}_m\left(1+R_i^{-1}\right)\left\{3\frac{A_m^{-2}+R_i}{1+R_i}\frac{\left(1+\alpha_m\varphi'\right)^2}{3-\varphi^{'2}}-1\right\}
\end{equation}
showing that the fictive DE tracks the cosmological evolution of the pressureless matter components. Therefore, this model mimics the action of an additional fluid with negative pressures although the matter content contains exclusively pressureless matter (ordinary matter and DM). In this context, the cosmic acceleration appears only for visible matter observers, it is a manifestation of SEP violation resulting from a relaxation of the WEP. The energy density of this fictive exotic DE fluid is proportional to the energy densities of both ordinary matter and AWE. Therefore, DE becomes a cosmological mechanism related to the rising influence of pressureless matter in the Universe, which explains why cosmic coincidence occured only recently.\\
\\
To be more accurate, we need to characterize the early cosmological evolution of this fictive exotic DE fluid, when
the scalar field starts deviating from GR
at the beginning of the matter-dominated era. Without loss of generality, we can consider for the effective coupling function
$\mathcal{A}(\varphi)$ the mexican-hat shape that is specified by the parameterization (\ref{couplings}). During radiation-dominated era, the scalar field has been damped by cosmic expansion
and pushed toward the minimum of $A_m(\varphi)$ that corresponds to GR (cf. \cite{convts,awedm}). 
However, as the universe progressively enters matter-dominated era $\omega\rightarrow 0$, this point where $\alpha_m(\varphi)=0$ becomes unstable and the scalar field starts slowly
rolling toward the attractor (\ref{phi_inf}). Therefore, we can simplify Eq. (\ref{dyn_2}) by assuming $\varphi^{'2}\ll 3$, $\omega= 0$ and linearizing $\mathcal{A}(\varphi)$ around $\varphi=0$ (see also \cite{awedm}) to get:
\begin{equation}
\label{earlyphi}
2\varphi''+3\varphi'+3 K_0 \varphi=0
\end{equation}
where $K_0=d\aleph(\varphi)/d\varphi|_{\varphi=0}$ is given by (\ref{dalephdphi}):
\begin{equation}
K_0=k\frac{R_i-R_\infty}{1+R_i},
\end{equation}
which is negative, resulting in the initial repulsion from GR. 
The general solution to (\ref{earlyphi}) is therefore
$$
\varphi(\lambda)=A_+e^{\nu_+ \lambda}+A_- e^{\nu_- \lambda},
$$
with
\begin{equation}
\nu_\pm=\frac{3}{4}\left(-1\pm\sqrt{1-\frac{8}{3}K_0}\right)\cdot
\end{equation}
The decaying mode in $\exp(\nu_-\lambda)$ can be neglected so that $\varphi$ behaves as a power-law of the bare scale factor $a$:
\begin{equation}
\label{phi_an}
\varphi(a)\approx \varphi_i\left(\frac{a}{a_i}\right)^{\nu}
\end{equation}
where $\varphi_i\approx 0$ is the initial condition at the end of radiation-dominated era $a=a_i\approx 10^{-3}$ and where we set $\nu=\nu_+$.
Using Eq. (\ref{phi_an}), the density of the fictive DE fluid (\ref{eff}) can be approximated by
$$
\tilde{\rho}_{DE}\approx 2 \tilde{\rho}_m (1+R_i^{-1}) k \varphi_i^2 \nu \left(\frac{\tilde{a}}{a_i}\right)^{2\nu}
$$
as we can consider that $A_m(\varphi)\approx 1$ because $\varphi$ stays small when the approximation (\ref{earlyphi}) holds.
This fictive DE fluid therefore mimics a cosmological constant when $\nu=3/2$ (as $\rho_m\approx \tilde{a}^{-3}$) and raises for $\nu>3/2$.
In general, the cosmological constant $\Lambda$ is a good approximation of the late cosmological evolution of the fictive DE fluid (see also Figures 3 \& 5).
As illustrated in Figure 5, while the radiation and baryon energy densities decrease in $\tilde{a}^{-4}$ and $\tilde{a}^{-3}$, respectively, like in GR as they are ruled by a restricted WEP, DM decreases faster than $\tilde{a}^{-3}$ shortly before today ($\tilde{a}=1$) because of the decrease
of the DM inertial mass. The cosmic gravitational strength (\ref{gc}) consequently increases which can be represented by a fictive DE density growing as a power-law of the scale factor. This DE slowly freezes after DE domination over baryons at $\tilde{a}\approx 0.5$, therefore mimicking a cosmological constant (horizontal line in Figure 5).
It is also important to note that the mechanism is not affected by the phase transition where the AWE particles become non-relativistic ($\omega_{awe}=1/3$, located arbitrarily at $\tilde{a}\approx 10^{-6}$ in Figure 5).\\
\\
Cosmic coincidence is achieved in (\ref{eff}) when $\rho_{DE}\approx\rho_m (1+R_i^{-1})$, i.e. when $\alpha_m(\varphi) \varphi'\approx \sqrt{2}-1$,
or equivalently
\begin{equation}
\label{coinc}
k \varphi_i^2\nu \left(\frac{\tilde{a}_c}{a_i}\right)^{2\nu}\approx 0.4142\cdot
\end{equation}
Using this approximation, we can study the sensitivity of cosmic coincidence to the model parameters. In Figure 6, we illustrate the values of the scalar coupling strength $k$ required to ensure 
a cosmic coincidence occuring at $\tilde{a}_c=0.5$ in (\ref{coinc}), for given values of the relative concentrations of baryons and DM $R_i$ and initial departure from GR $\varphi_i$, both at 
CMB epoch\footnote{In Figure 6, the ratio $R_\infty$ between the scalar charges $\alpha_m$ and $\alpha_{awe}$ has been set to unity (see also section 6).} With a scalar coupling strength $k$ varying from $1$ to $100$, it is possible to explain coincidence from an impressively wide range of couples in the plane $(R_i,\varphi_i)$.
Such values of the scalar coupling strength are very natural for
non-minimally coupled theories as this simply translates that the coupling strength $k$ to the gravitational scalar $\varphi$ is roughly of the same amplitude than the coupling strength $\kappa=8\pi$ to its tensorial counterpart $g_{\mu\nu}$.
Cosmic coincidence therefore appears naturally from the typical amplitude of the model parameters: $R_i\approx 1$, $\varphi\ll 1$ and $k\approx 1$.
 We see in Figure 6 that the approximation (\ref{coinc}) also puts a crude upper limit on the departure from GR at the end of radiative-dominated era: $\varphi_i< 10^{-4}$ (for
$R_i\approx 0.2$ and $\nu=3/2$), or in terms of the observed gravitational strength (\ref{gc}): $G_c/G-1\le 10^{-8}$. The AWE mechanism therefore does not require a substantial departure from GR at CMB to do its job. On the contrary, the closeness required for cosmic coincidence that is obtained from our approximation here is in agreement with other cosmological constraints in the radiative era, for instance on BBN \cite{BBN2,BBN3}.
\section{A Mach-Dirac Principle for gravitation}
The mass-variation of the AWE, which consists on a violation of WEP between visible and invisible sectors, 
implies a compulsory violation of the SEP. This reasoning on the equivalence principles links together two old but crucial ideas in gravitational physics: Mach's principle and Dirac's hypothesis. Mach's principle \cite{bd} was one of Einstein's main inspiration \cite{pais} when he conceived GR and posits that inertia and acceleration are relative and should therefore be refered to distant matter. In particular, Mach advanced that inertial masses could only be defined with respect to the entire matter distribution in the Universe. The other key idea is the hypothesis of varying observed gravitational strength $G_c$ (\ref{gc}) that Dirac formulated \cite{dirac} in order to justify some puzzling numerical coincidences between cosmological and quantum numbers. Here we show that the AWE hypothesis can merge into a \textit{Mach-Dirac principle} these thoughtful ideas by identifying the machian mass variation of AWE as the source of Dirac's varying gravitational strength. \\
\\
Indeed, a new remarkable feature of this model is that, besides of its cosmological predictions, it gives a hint on a new relation between microphysics and gravitation.
This is expressed in terms of the gravitational strength $G_{c}$ (\ref{gc}) and DM mass $m_{{DM}}\equiv \bar{m}_{DM} M(\Phi)$, where $\Phi(x^\mu)$ is the dimensionless space-time dependent
coupling, $G_{N}$ is Newton's constant and $\bar{m}_{DM}$ is the DM mass that would be measured in Earth-based laboratories.
To obtain the cosmological evolution described above (see also Figure 2), we have seen that the gravitational coupling $G_{c}$ and the DM mass-scaling must be inversely proportional (see Eq.(\ref{as}) and \cite{awedm}), which also means that ordinary matter and AWE must have opposite scalar charges.
In terms of the dimensionless coupling $\Phi$, we therefore have:
\begin{equation}
M(\Phi)=\Phi^{\frac{1+R_\infty}{2}}
\end{equation}
where $R_\infty$ is also the ratio $\rho_{m}/\rho_{awe}(t\rightarrow\infty)$ at which baryons and DM densities freeze into the above-mentionned attractor (see Figure 2).
From the cosmological data on supernovae, we find $R_\infty=1.23^{+0.96}_{-0.67}$ ($R_\infty=1.35^{+1.11}_{-0.85}$) at the 68\% confidence level for HST (SNLS) data sample. The measured value of $R_\infty$ close to unity indicates that the scalar charges of ordinary matter and DM are \textit{exactly} opposite. 
Indeed, $R_\infty=1$ implies that $\alpha_m(\varphi)=-\alpha_{awe}(\varphi)$ ($A_m(\varphi)=A_{awe}^{-1}(\varphi)$ ; $M(\Phi)=\Phi$) which constitutes
a hint to a possible new symmetry between the hidden and visible sectors. In other words, this value of $R_\infty$ also points to an intriguing relation between the constant mass of baryons $m_b$ and the changing DM $m_{DM}$ and observed gravitational strength $G_{c}$ (\ref{gc}):
\begin{equation}
G_c(x^\mu)\; m_{b}\;m_{DM}(x^\mu)=G_{N}\; m_{b}\;\bar{m}_{DM},
\label{md}
\end{equation}
where the bar means the Earth laboratory value. 
This dimensionless relation is frame-independent and imposing $R_\infty=1$ is the only way of constructing a constant ratio
between the masses of the particles from the visible and invisible sectors and the Planck scale. Although this relation does not fix the bare mass of DM, it rules its scaling by imposing a conservation of the product of the gravitational \textit{charges} of baryons and DM. 
This important phenomenological law, directly deduced from cosmological data and linking together gravitational scales and masses of visible and invisible matter glimpses at the intimate nature of gravitation. The deep meaning of Eq.(\ref{md}) constitutes a crucial question for the many fundamental approaches that aim to unify gravity and microphysics with
explicit space-time dependancies of masses and couplings \cite{ashtekar,connes,gasperini1,gasperini2}.
\section{Conclusion and perspectives}
Let us first review the underlying assumptions on which this paper is based in an exhaustive way.
\begin{itemize}
\item[(a)] The invisible sector is \textit{abnormally weighting}, i.e. it does not experience the background spacetime with the same gravitational strength than ordinary visible matter (see Eq.(\ref{s_awe2})). 
For visible matter observers, the particles of the dark sector therefore appear to carry a space-time dependent mass that modifies the gravitational coupling strength (see Eq.(\ref{s_awe})). 
To describe this relaxation of the WEP,  it is necessary to extend the usual tensor-scalar theories of gravitation \cite{ts, bd} in which a violation of SEP only (varying gravitational coupling) is considered. 
In these approaches, gravitation is mediated by both a metric field and a scalar companion describing the variation of the gravitational strength while an arbitrary function describes
the dynamics of the gravitational scalar  (the scalar coupling function to matter $A_m(\varphi)$ in (\ref{s_awe2}) or the Brans-Dicke function $\omega_{BD}(\Phi)$ in (\ref{s_awe})). To account for the WEP violation brought
by the dark sector, an additional function has to be introduced (the scalar coupling function to AWE  $A_{awe}(\varphi)$ in (\ref{s_awe2}) or the mass scaling of AWE $M(\Phi)$ in (\ref{s_awe})).
\\
\item[(b)] In order to ensure a large-scale departure from GR that provides cosmic acceleration, the scalar charges $\alpha_m$ and $\alpha_{awe}$ of the visible and invisible sectors must be of opposite signs (see (\ref{kg}) and (\ref{kg_awe})). This means that the coupling between the scalar and matter or AWE,
$A_m(\varphi)$ and $A_{awe}(\varphi)$, are reciprocal functions. Therefore, we are left with specifying the single remaining matter coupling function $A_m(\varphi)$. The very general choice
of a matter coupling function $A_m(\varphi)$ with a global minimum ensures both  local agreement with constraints on the SEP and a modified cosmological convergence mechanism toward 
a GR-like state (see section 3 for details). This mechanism is as follows: convergence toward GR during radiation-dominated era (standard TST), departure from GR at the beginning of matter-dominated era (transient cosmic acceleration phase) and
finally re-convergence toward a GR-like gravitation with a different gravitational coupling on large-scales.
\\
\item[(c)] Doing so, the model is characterized by four parameters: $R_i=\rho_m/\rho_{awe}(a_i)$ the relative amount of visible and invisible matter at CMB epoch, $R_\infty=|\alpha_{awe}/\alpha_m|$ the 
(absolute value of the) ratio between the scalar charges
of invisible and visible matter, $\varphi_i$ the initial departure from GR at CMB epoch and $k$ the scalar coupling strength. None of these parameters need to be fine-tuned to reproduce cosmological data (see Figure 6).
A natural choice of the scalar coupling strength $k$ is to take it of order unity, so that
matter couples to the metric and its scalar companion with roughly the same strength. Finally, we remind that $R_i$ is of course not a new free parameter as it is already present in the cosmic concordance.
\\
\end{itemize}
As formal results, we have therefore generalized TST to a WEP relaxation, studied the complete revisited convergence mechanism toward GR in the presence of an AWE component and established its link with cosmic acceleration. We can now take back the AWE-full problems of cosmic concordance presented in the introduction to see what improvements are brought by the results of this paper.
\begin{itemize}
\item[(i)] \textbf{Nature of DM and DE}\\
Cosmic acceleration (DE) is obtained  from the anomalous gravitational properties of DM. DE is no more an additional exotic component that is physically unrelated to matter. 
The space-time dependance of the DM mass, a property due to direct interactions of DM with one or more extra fields, creates a long-ranged gravitational force mediated by (massless here) scalars that exerts between the opposite scalar charges of the visible and invisible sectors. DE is now identified to the dynamics of gravitational scalar field that causes the observed cosmic acceleration. This dynamics that is essentially active on large scales where the amounts of ordinary and invisible matter are similar, leaving GR unchanged 
where DM is sub-dominant. \\
\item[(ii)] \textbf{Negative pressures}\\
Cosmic acceleration arises from the gravitational feedback of DM mass-variation without requiring to any explicit negative pressures. Ordinary 
and invisible matter undergo a long-ranged scalar interaction but they are both pressureless on cosmological scales. Cosmic acceleration is nothing but an observable consequence of a modification of gravity (WEP relaxation and SEP violation) on large-scales. If we translate this mechanism in terms of a standard FLRW expansion with exotic fluids of negative pressures, then the effective DE fluid appears to have a phantom equation of state $\omega<-1$
(see \cite{awedm}). \\
\item[(iii)] \textbf{Cosmic coincidence}\\
Cosmic acceleration is intimately related to invisible non-relativistic matter, and can only appear during matter-dominated era, after a typical time scale of a few efoldings. This time-scale is related to the quite natural assumptions on the model parameters (see point (c) above and section 5). The coincidence on DE is therefore explained in terms of the one between
baryons and DM ($R_i\approx 1$) .
\item[(iv)] \textbf{Fate of the Universe}\\
Although the effective DE fluid can mimic a phantom equation of state for some time, cosmic acceleration phases are transient. The asymptotic state is Einstein-de Sitter (matter-dominated) cosmology
when the gravitational coupling has finally reached an equilibrium where visible and invisible matter equally \textit{weigh}. The Big Rip and eternal cosmic acceleration denouements are avoided.

\end{itemize}
In addition, we would like to emphasize two corollary results of great relevance. The first is that
our model allows to successfully measure, independently of the CMB and BBN, the amount of baryons in the Universe from the Hubble diagram cosmological test \textit{alone}, which is not possible in cosmology based on GR. Secondly, the AWE hypothesis, and more generally any approach where DE couples directly to matter, leads to supplementary observational tests than minimally coupled DE like the cosmological constant. Indeed, the cosmological constant only affects the background expansion and does not cluster. It can therefore only be tested through the cosmic expansion and the linear growing rate of density fluctuations, making it very costly to test through the measurement of the equation of state and its constancy, for instance. More generally, minimally coupled models like quintessence are subject to the same restrictions at the exception that quintessence undergoes slight gravitational collapse, therefore leaving additional fine imprints on structure formation. With non-minimally coupled models, like this one or more generally interacting DM and DE, it is possible to build laboratory tests  \cite{chameleon2} of the nature of both of them as well as tests of modified gravitational physics on large-scales \cite{kamionkowski,bertolami,lius} where DM is profuse. And if such tests were finally found inconclusive, they would therefore constitute independent proofs of $\Lambda\textrm{CDM}$... 
\\
\\
Arising from a critical discussion on the equivalence principles, the AWE Hypothesis leads to a new elegant theory of gravitation
for dealing with the invisible sector, \textit{tensor-scalar anomalous gravity}, that glimpses beyond GR.
Applying this theory to cosmology allowed us to successfully explains \textit{cosmic acceleration} in terms of the anomalous gravity arising as a feedback of \textit{dark matter} mass variation. Doing so, this suppresses the need of an independent additional DE component with an almost intractable coincidental dominance.
But this interpretation of DE also offers to physicists new observational and theoretical challenges.
The existence of an abnormally weighting type of energy, that one can identify to DM from the results presented here, not only affects the background expansion, at the opposite of minimally coupled models of DE, it also modifies gravitational physics on large-scales where DM is profuse. 
This offers new perspectives to test the nature of DM, and indirectly of DE itself. 
The theoretical challenge lies in explaining why the gravitational energy scale, i.e. the Planck mass, would change
with the inertial mass of DM particles, as dictated by Eq. (\ref{md}). This Mach-Dirac principle is intimately linked to an enticing new symmetry between the visible and invisible sectors: the exact oppositeness of their scalar charges. The existence of such a long-ranged fifth force mediated by a scalar exchanged between the opposite charges of the visible and invisible sectors, can be hoped to be established in local experiments \cite{kaloper,chameleon2}. 
If the AWE hypothesis really constitutes the \textit{missing link} between dark matter and dark energy, then elucidating the intimate nature of gravitation could be closer than ever. 
\section*{Acknowledgements}
A.F. is supported by the Belgian "Fonds de la Recherche Scientifique" (F.N.R.S. Postdoctoral Researcher) and associated researcher at CP3 (UCL, Belgium).
Numerical simulations were partly made at UCL HPC Center (Belgium) under project FRFC N°2.4502.05, and the authors warmly thank V. Boucher for technical help.
The authors are very grateful to P.-S. Corasaniti and C. Ringeval for interesting discussions and careful reading of this paper.
\newpage
\begin{figure}
\begin{center}
\includegraphics[scale=0.4]{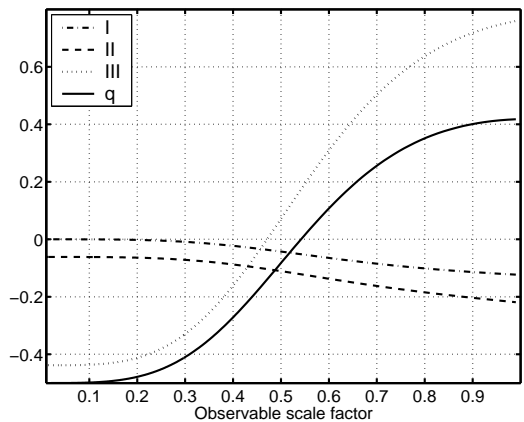}
\end{center}
\caption{Observable acceleration factor $q=d^2\tilde{a}/d\tilde{t}^2/(\tilde{a}\tilde{H}^2)$ obtained from the fit to SNe Ia data and its constitutive components from Eq. (\ref{acc_obs}) ($R_\infty=1$, $R_i=0.14$, $k=5.42$, $\varphi_i\approx 10^{-9}$, $\Omega_m^0=0.05$, $\Omega_{\textrm{awe}}^0=0.24$,
$\chi^2/\textrm{dof}=1.06$ on SNLS data set, $\chi^2/\textrm{dof}$($\Lambda$CDM)=$1.05$) } \label{fig1}
\end{figure}

\begin{figure}
\begin{tabular}{cc}
\includegraphics[scale=0.4]{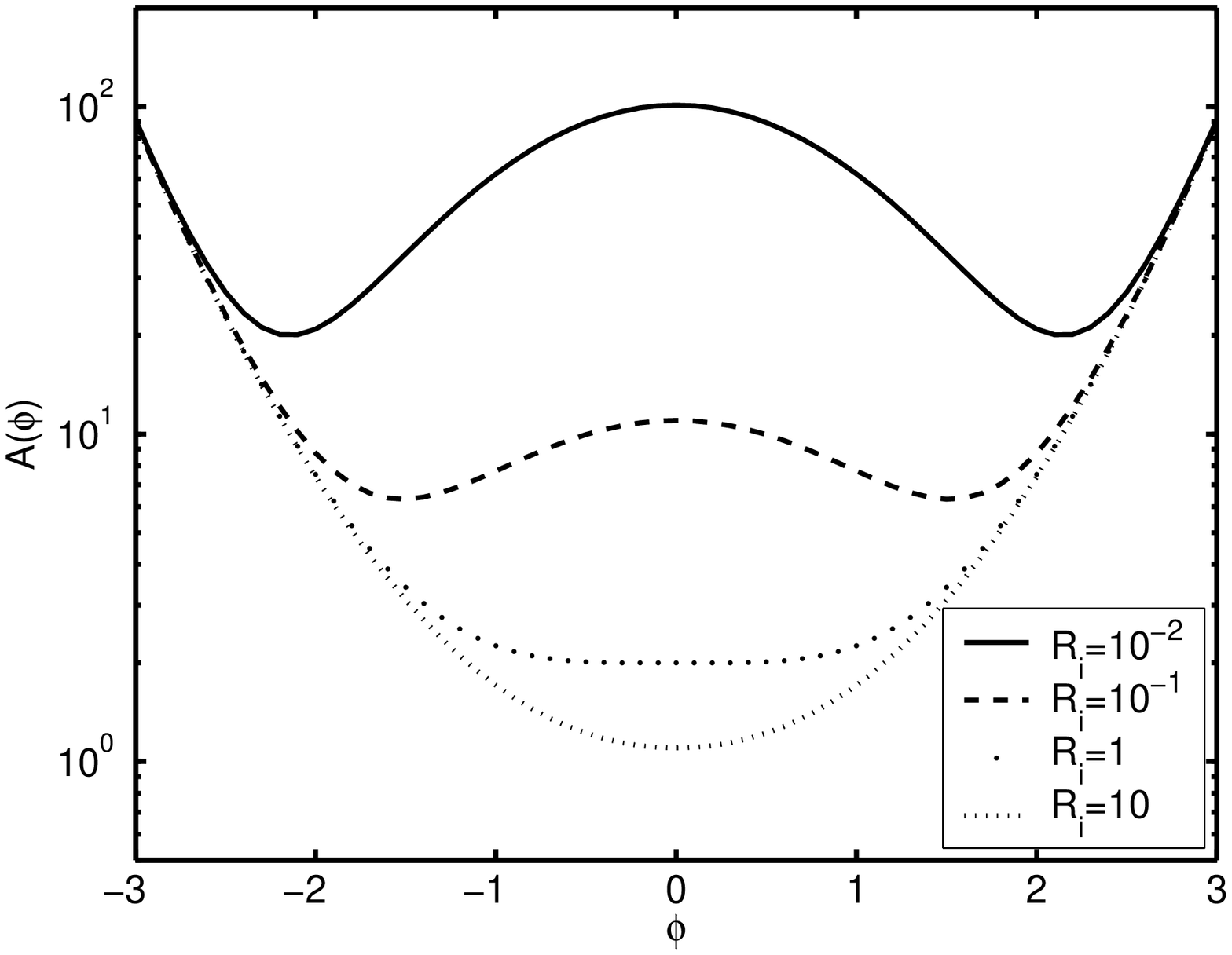} &
\includegraphics[scale=0.4]{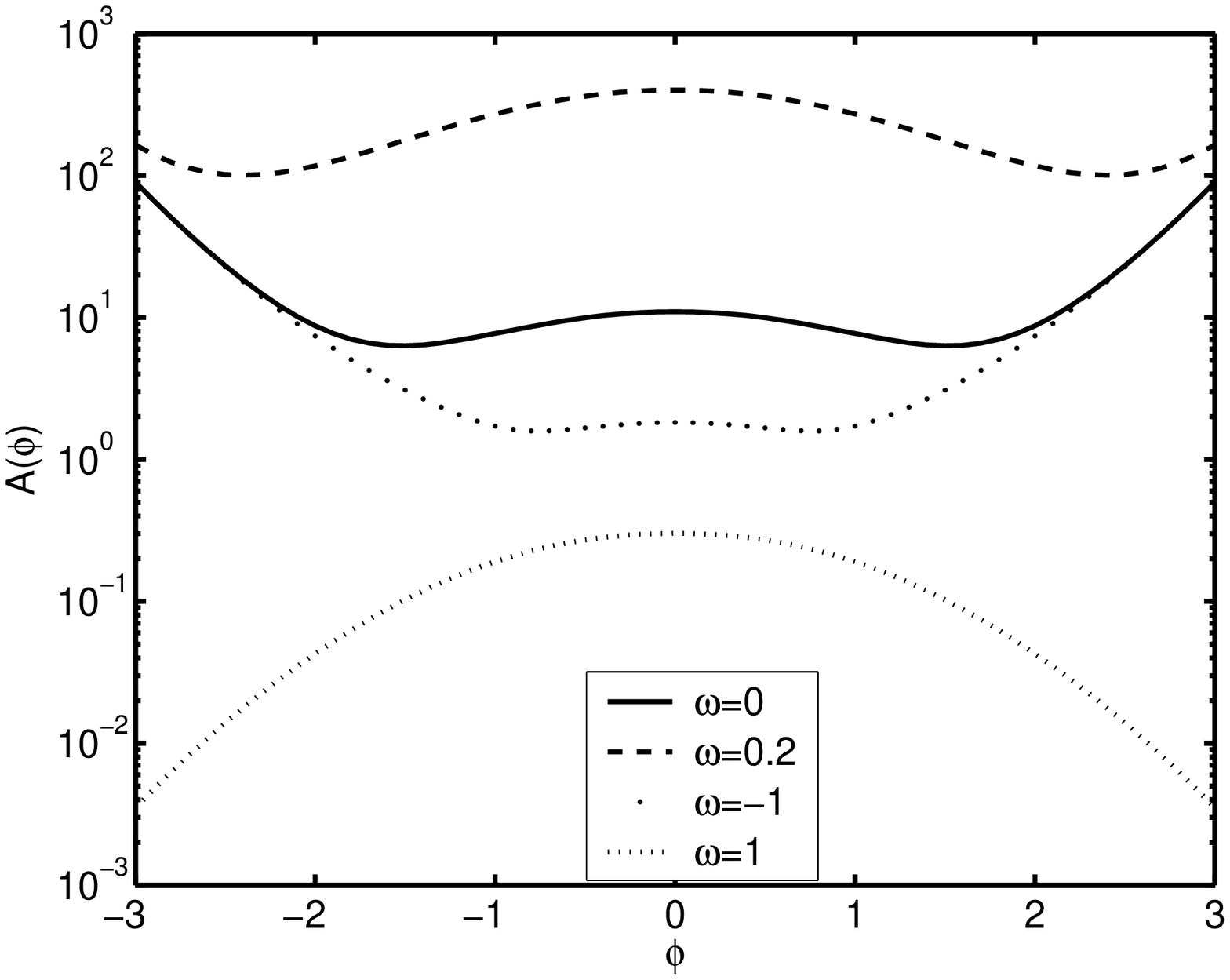} 
\end{tabular}
\caption{Effective potential for the dynamics of the gravitational coupling in the matter-dominated era (left panel) and for different
values of the equation of state parameter $\omega$ (right panel)  \label{fig2}}
\end{figure}

\begin{figure}
\begin{tabular}{cc}
\includegraphics[scale=0.3]{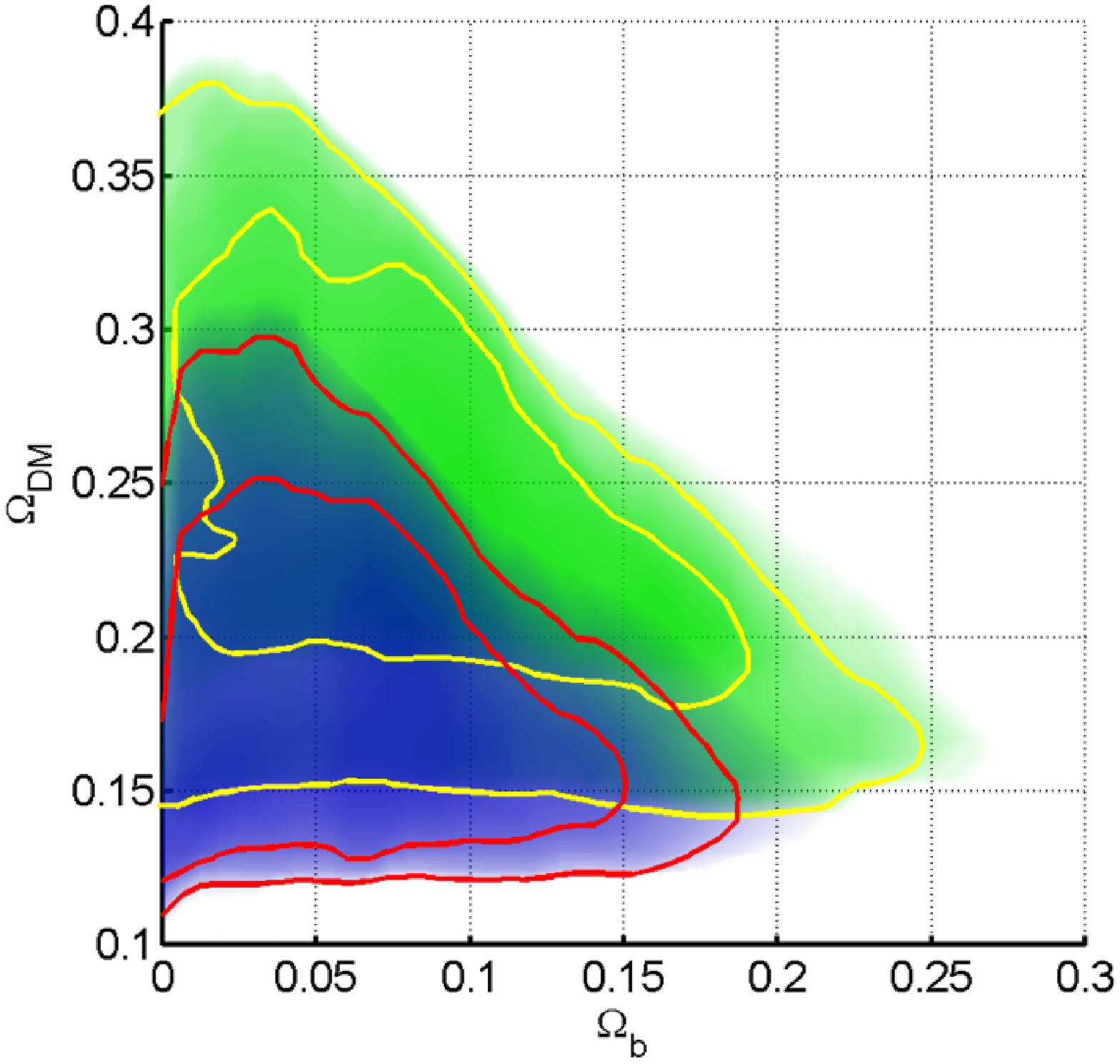} &
\includegraphics[scale=0.3]{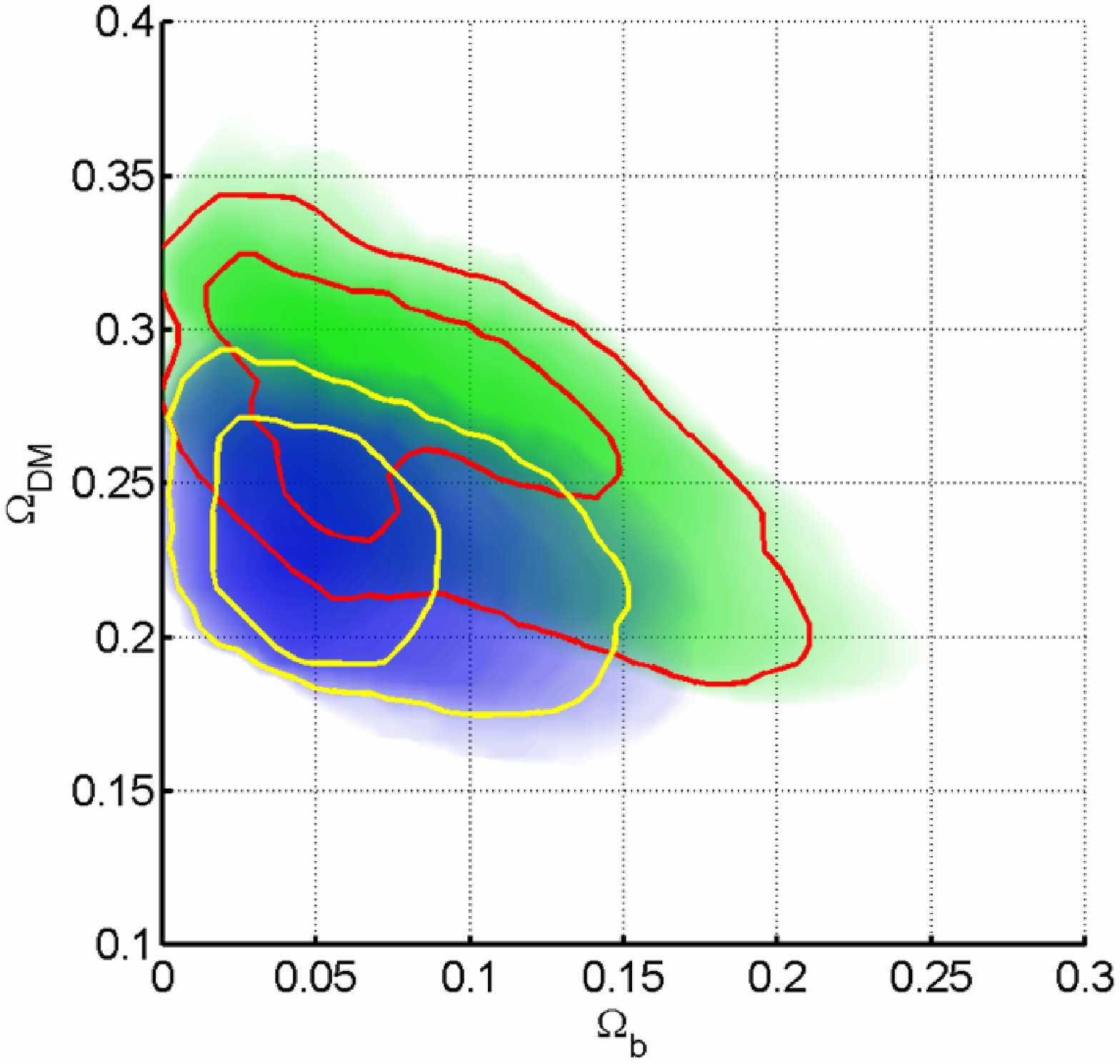}
\end{tabular}
\caption{68\% (inner) and 95\% (outer) confidence contours of the density parameters associated to ordinary and AWE (dark matter) ($\Omega_b$ and $\Omega_{DM}$, respectively) 
for HST (green) and SNLS (blue) supernovae data sets. Left panel is for the parameterization (\ref{couplings}) while right panel is for parameterization (\ref{couplings2}),
illustrating the robustness of the predictions when the shape of Figure 2 is preserved \label{fig3}}
\end{figure}

\begin{figure}
\begin{center}
\includegraphics[scale=0.4]{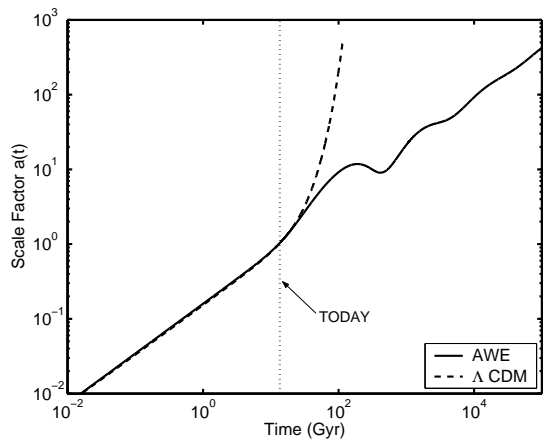}
\end{center}
\caption{Comparison between the cosmic histories predicted by the AWE model in the Dicke-Jordan frame and the concordance model $\Lambda\textrm{CDM}$.\label{fig4}}
\end{figure}

\begin{figure}
\begin{center}
\includegraphics[scale=0.4]{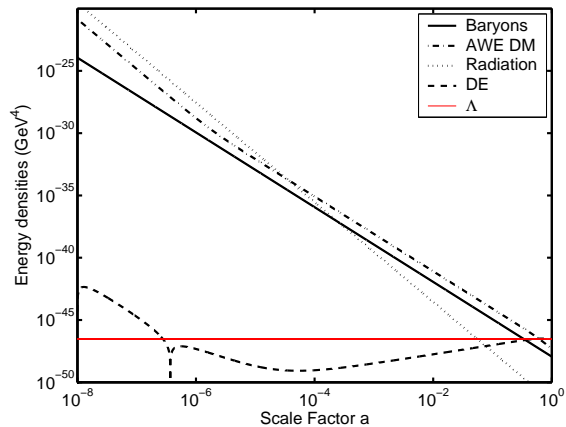}
\end{center}
\caption{Evolution of the observable energy densities of the cosmic components ordinary matter, AWE and scalar field $\varphi$   \label{fig5}}
\end{figure}

\begin{figure}
\begin{center}
\includegraphics[scale=0.4]{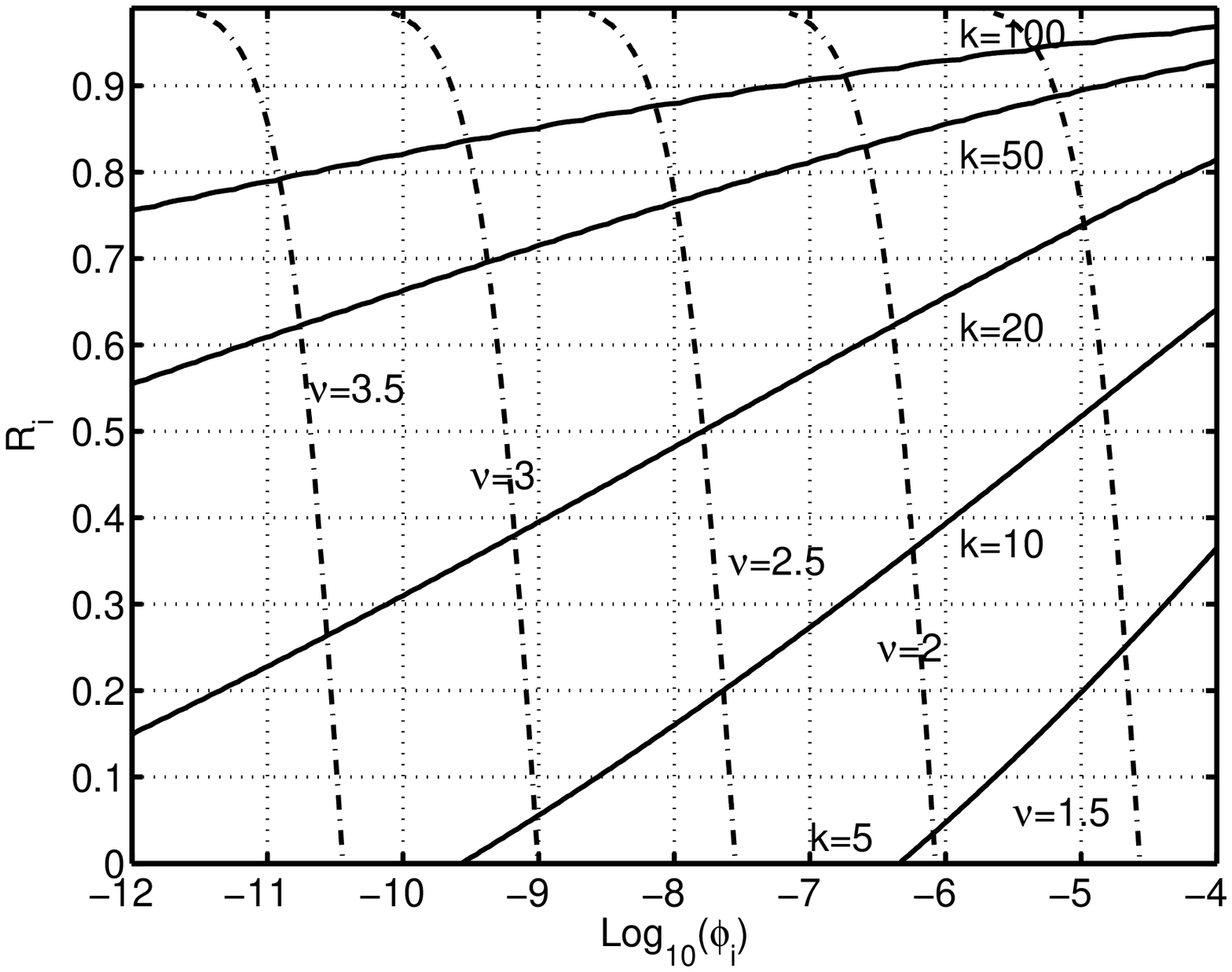}
\end{center}
\caption{Required values of the scalar coupling strength $k$ (solid lines), and consequent values of $\nu$ (dash-dotted lines) to ensure a cosmic coincidence occuring at $\tilde{a}=0.5$, as a function of the initial conditions $R_i$ and $\varphi_i$  at CMB epoch  } \label{fig6}
\end{figure}


\section*{References}
\begin{thebibliography}{30}
\bibitem{wmap1} D.N. Spergel et. al., Astrophys. J. Suppl. \textbf{148} 1-27 (2003).
\bibitem{wmap3} D.N. Spergel et. al., Astrophys. J. Suppl. \textbf{170} 377 (2007).
\bibitem{wmap5} E. Komatsu et al.,  arXiv:0803.0547 (2008)
\bibitem{perlmutter} S. Perlmutter et al., Astrophys.J. 517 (1999) 565-586.
\bibitem{riess} A.G. Riess et al., Astrophys. J. \textbf{607} 665-687 (2004).
\bibitem{snls} P. Astier et al., Astron. Astrophys. \textbf{447} 1 31-48 (2006).
\bibitem{einstein} A. Einstein, "\textit{Cosmological Considerations on the General Theory of Relativity}", in "\textit{The Principle of Relativity}", Dover, 1952.
\bibitem{weinberg} S. Weinberg, Rev. Mod. Phys. \textbf{61} 1-23 (1989).
\bibitem{copeland} E.J. Copeland, M. Sami \& S. Tsujikawa, Int.J.Mod.Phys. D15 1753-1936 (2006).
\bibitem{brax} P. Brax \& J. Martin, JCAP 11, 008 (2006).
\bibitem{amendola} L. Amendola, Phys. Rev D \textbf{62} 043511 (2000).
\bibitem{farrar} G.R. Farrar \& P.J.E. Peebles, Astrophys. J. \textbf{604} 1-11 (2004).
\bibitem{corasaniti} S. Das, P.-S. Corasaniti \& J. Khoury, Phys.Rev. D73 (2006) 083509.
\bibitem{will} C.M. Will, Liv. Rev. Rel. \textbf{9}, 3 (2006).
\bibitem{su} Y. Su et al. Phys. Rev. D 50, 3614 - 3636 (1994)
\bibitem{kamionkowski} M. Kesden \& M. Kamionkowski, Phys.Rev. D74 (2006) 083007;
Phys.Rev.Lett. 97 (2006) 131303.
\bibitem{massd} C.T. Will \& G.C. Ross, Nucl. Phys. B 311, 253 (1988);\\
J. Ellis, S. Kalara, K.A. Olive \& C. Wetterich, Phys. Lett. B 228, 264 (1989);\\
C. Wetterich, Astron. Astrophys. 301, 321 (1995);\\
G.W. Anderson \& S.M. Caroll, astro-ph/9711288;\\
G. Huey, P.J. Steinhardt, B.A. Ovrut \& D. Waldram, Phys. Lett. B 476, 379 (2000);\\
D. F. Mota, J. D. Barrow, Mon. Not. Roy. Astron. Soc. 349, 291 (2004), astro-ph/0309273\\
D. F. Mota, J. D. Barrow, Phys.Lett.B 581 141-146 (2004), astro-ph/0306047\\
A. Ringwald \& L. Schrempp, JCAP 0610, 012 (2006)\\
A. W. Brookfield  et al., PRD 73, 083515 (2006)\\
R. Fardon et al., JCAP 0410, 005 (2004)\\
L. Amendola, Phys.Rev. D62 (2000) 043511, \\ 
S. Das, P.-S. Corasaniti \& J. Khoury, Phys.Rev. D73 (2006) 083509
\bibitem{caroll} S.M. Carroll, Spacetime and Geometry: An Introduction
to General Relativity, San Francisco, Addison-Wesley (2004)
\bibitem{bigrip} R.R. Caldwell, M. Kamionkowski, N. Weinberg, Phys. Rev. Lett. 91, 071301 (2003)
\bibitem{ts} P. Jordan, Nature 164, 637 (1949)\\
M. Fierz, Helv. Phys. Acta 29, 128 (1956)
\bibitem{bd} C. Brans \& R.H. Dicke, Phys. Rev. \textbf{124}, 925-935 (1961).
\bibitem{convts} A. Serna, J.-M. Alimi and A. Navarro, Class.Quant.Grav. 19 (2002) 857-874;\\
A. Serna and J.-M. Alimi, Phys.Rev. D53, 3074 (1996);\\
T. Damour and K. Nordtvedt, Phys. Rev. D48 (8),
  3436-3450 (1993);\\
T. Damour and K. Nordtvedt, Phys. Rev. Lett. 70 (15),
  2217-2219 (1993)
\bibitem{barrow} J.D. Barrow, Phys. Rev. D 47, 5329 - 5335 (1993)
\bibitem{damour2} T. Damour, D. Polyakov, Nucl.Phys. B423 (1994) 532-558.
\bibitem{awebi} A. F\"uzfa \& J.-M. Alimi, Phys. Rev. D \textbf{73} 023520 (2006).
\bibitem{awe} A. F\"uzfa \& J.-M. Alimi, Phys. Rev. Lett. \textbf{97} 061301 (2006)
\bibitem{awedm} A. F\"uzfa \& J.-M. Alimi, Phys. Rev. D \textbf{75} 123007 (2007).
\bibitem{awedm2} J.-M. Alimi \& A. F\"uzfa, Int.J. Mod. Phys. D 16, 2587 - 2592 (2007).
\bibitem{damour} T. Damour, G.W. Gibbons \& C. Gundlach, Phys. Rev. Lett. \textbf{64}, 123 (1990).
\bibitem{catena} R. Catena, M. Pietroni \& L. Scarabello, Phys.Rev. D70 (2004) 103526.
\bibitem{chameleon} J. Khoury, A. Weltman, Phys. Rev. Lett. 93, 171104 (2004);\\
J. Khoury, A. Weltman, Phys. Rev. D69, 044026 (2004);\\ P. Brax, C. van de Bruck, A.C. Davis, J. Khoury, A. Weltman, Phys. Rev. D 70, 123518
(2004)\\
D. F. Mota \& D. J. Shaw, Phys. Rev. Lett. 97 151102 (2006);\\
D. F. Mota \& D. J. Shaw, Phys. Rev. D 75, 063501 (2007), hep-ph/0608078
\bibitem{kaloper} N. Kaloper, Phys.Lett.B653, 109-115 (2007).
\bibitem{gef2} T. Damour \& G. Esposito-Farese, Phys. Rev. Lett. 70, 2220 (1993).
\bibitem{gef} T. Damour \& G. Esposito-Farese, Phys.Rev. D58 (1998) 042001;
G. Esposito-Farese, Proceedings of the 10th Marcel Grossmann Conference, MG10, Rio de Janeiro (Brazil), World Scientific (2005), 647-666, gr-qc/0402007. 
G. Esposito-Farese, Proceedings of the 9th Marcel Grossmann Conference, MG9, Roma (Italy), World Scientific (2002), 2041, gr-qc/0011114. 
\bibitem{BBN} R.H. Cyburt, B.D. Fields, \& K.A. Olive, Phys. Lett. B, 567 (3), 227-234 (2003).
\bibitem{sahni} U. Alam, V. Sahni, T. D. Saini \& A. A. Starobinsky, Mon.Not.Roy.Astron.Soc. 344 (2003) 1057;
S. Capozziello, V.F. Cardone \& V. Salzano, arXiv:0802.1583
\bibitem{BBN2} A. Serna, J.-M. Alimi \& A. Navarro, Class. Quant. Grav. 19 (2002) 857-874.
\bibitem{BBN3}  R.H. Cyburt, Phys. Rev. D 70, 023505 (2004).
\bibitem{pais} A. Pais, \textit{Subtle Is the Lord: The Science and the Life of Albert Einstein}, Oxford University Press, 1982.
\bibitem{dirac} P.A.M. Dirac, Proc. Roy. Soc. A \textbf{165}, 199-208 (1938).
\bibitem{ashtekar} A. Ashtekar, \textit{Gravity, geometry and the quantum}, appeared in Albert Einstein Century International Conference, J.-M. Alimi \& A. F\"uzfa Eds, AIP Conference Proceedings, 861 (2006).
\bibitem{connes} A. Connes, \textit{Noncommutative geometry and physics}, appeared in Albert Einstein Century International Conference, J.-M. Alimi \& A. F\"uzfa Eds, AIP Conference Proceedings, 861 (2006).
\bibitem{gasperini1} M. Gasperini, Lect. Notes Phys. 737, 789 (2007).
\bibitem{gasperini2} M. Gasperini, \textit{Elements of string cosmology}, Cambridge University Press, Cambridge (2007).
\bibitem{chameleon2} P. Brax, et al. Phys. Rev. D 76, 085010 (2007), Phys. Rev. D 76, 124034 (2007) 
Ph. Brax, C. van de Bruck, A.-C. Davis, Phys. Rev. Lett. 99, 121103 (2007)
\bibitem{bertolami}  O. Bertolami, F. Gil Pedro, \& M. Le Delliou, Phys. Lett. B 654 (5), 165-169 (2007).
\bibitem{lius} Liu, D. \& Liu W. Phys. Rev. D 77 027301 (2008).
\end{thebibliography}
\end{document}